# The Brightness Temperature of the Quiet Solar Chromosphere at 2.6 mm

**Kazumasa Iwai**[1,*]. Masumi Shimojo[2]. Shinichiro Asayama[2].
Tetsuhiro Minamidani[3]. Stephen White[4]. Timothy Bastian[5]. and Masao Saito[3]

[1]National Institute of Information and Communications Technology, Koganei 184-8795, Tokyo, Japan
[2]Chile Observatory, National Astronomical Observatory of Japan, Mitaka, Tokyo 181-8588, Japan
[3]Nobeyama Radio Observatory, National Astronomical Observatory of Japan, Minamimaki, Nagano 384-1305, Japan
[4]Space Vehicles Division, Air Force Research Laboratory, Albuquerque, NM, USA
[5]National Radio Astronomy Observatory, Charlottesville, VA, USA
*Corresponding author email: kazumasa.iwai@nict.go.jp

**Abstract**

The absolute brightness temperature of the Sun at millimeter wavelengths is an important diagnostic of the solar chromosphere. Because the Sun is so bright, measurement of this property usually involves the operation of telescopes under extreme conditions and requires a rigorous performance assessment of the telescope. In this study, we establish solar observation and calibration techniques at 2.6-mm wavelength for the Nobeyama 45-m telescope and derive the absolute solar brightness temperature accurately. We tune the superconductor–insulator–superconductor (SIS) receiver by inducing different bias voltages onto the SIS mixer to prevent saturation. Then, we examine the linearity of the receiver system by comparing outputs derived from different tuning conditions. Further, we measure the lunar filled beam efficiency of the telescope using the New Moon, and then derive the absolute brightness temperature of the Sun. The derived solar brightness temperature is 7700 ± 310 K at 115 GHz. The telescope beam pattern is modeled as a summation of three Gaussian functions and derived using the solar limb. The real shape of the Sun is determined via deconvolution of the beam pattern from the observed map. Such well-calibrated single-dish observations are important for high-resolution chromospheric studies because they provide the absolute temperature scale missing from interferometer observations.

**1. Introduction**

The brightness temperature of the Sun constitutes a basic property of the solar atmosphere. The main emission mechanism of the Sun at millimeter and sub-millimeter wavelengths is thermal free–free emission from the chromosphere, which is an atmospheric layer with temperature ranging between 6000 and 20,000 K. The opacity of thermal free–free emission depends on the temperature and density in the emission region. In addition, the Rayleigh-Jeans law is applicable to this wavelength range (Dulk, 1985); thus, we can develop vertical models of the chromospheric density and temperature distributions based on the brightness temperature spectrum of the thermal free-free emission.

A number of solar millimetric and sub-millimetric observations have been conducted (*e.g.*, Kosugi *et al*., 1986; Bastian *et al*., 1993; Lindsey and Kopp, 1995; Irimajiri *et al*., 1995; Lindsey *et al*.. 1995; White *et al*., 2006; Iwai and Shimojo, 2015; Iwai *et al*., 2016). It is known that millimetric thermal free–free emission is formed under local thermodynamic equilibrium (LTE) conditions. Further, analytic expressions exist that enable straightforward calculation of the brightness temperature expected from any model atmosphere. A large number of atmospheric modeling studies have been conducted taking advantage of this method, yielding various predictions for the millimetric and sub-millimetric radio emission of the Sun (*e.g.*, Zirin *et al*., 1991; Selhorst *et al*., 2005; Loukitcheva *et al*., 2014). Also, many new discoveries relating to the chromosphere have been reported based on data from the *Hinode* and *Interface Region Imaging Spectrograph* (IRIS) spacecraft, especially as regards chromospheric fine structures and dynamics (*e.g*., Okamoto and De Pontieu, 2011; De Pontieu *et al*., 2014). These high-resolution observations have led to the identification of complex chromospheric structures. On the other hand, the observed optical and ultraviolet (UV) lines are thought to be formed under non-LTE conditions. Hence, the observational results require non-LTE radiative transfer simulations to facilitate their interpretation. As a result, high-resolution observations of the chromosphere at millimetric and sub-millimetric wavelength are becoming increasingly important at present. Although interferometer observations can achieve higher spatial resolution, the interferometry technique filters out all angular scales greater than that measured by the shortest baseline. On the other hand, single-dish observations measure all angular scales from DC to the angular resolution of the antenna. Therefore, the single-dish observations allow us to measure the absolute brightness temperature accurately, and they can be used to supply the absolute temperature information missing from interferometer observations of smaller spatial scales.

However, large millimeter and sub-millimeter telescopes are not always able to observe the Sun. If the dish surface is too smooth, the intense infrared emission from the Sun enters the optical path and can damage telescope components, so many millimeter and sub-millimeter telescopes are prohibited from pointing near the Sun. Further, even though some telescopes can resist sunlight-generated heat, various technical difficulties exist as regards deriving the brightness temperature variations of the solar disc and its fine structures.

The purpose of this study is to establish observation and calibration sequences of the solar brightness temperatures at the millimeter wavelengths using single-dish telescopes. In this study, we demonstrate solar observation sequences at the millimeter wavelengths using the Nobeyama 45-m telescope. We extend the dynamic range of the receiver system using a "de-tuning method" explained below. Then, the linearity of the receiver system is examined using the Sun itself. We also model the beam pattern, including the broad sidelobes of the 45-m telescope, which have not previously been measured. These results, together with a beam efficiency estimation using the New Moon, enable us to derive the absolute solar brightness temperature and its fine structures. The methods to derive the brightness temperature of the Sun are described in Section 2. The data analysis and results are provided and discussed through comparison with previous results in Section 3. Lastly, we summarize the findings of this study in Section 4.

## 2. Methods
### 2.1 Observation
The Nobeyama 45-m-diameter radio telescope is a ground-based, millimeter-wavelength radio telescope operated by the Nobeyama Radio Observatory (NRO), which is part of the National Astronomical Observatory of Japan (NAOJ). The typical spatial resolution of the Nobeyama 45-m radio telescope is 15´´ at 115 GHz. In this study, we employed a

superconductor–insulator–superconductor (SIS) receiver (S100) whose bias voltage and local Gunn diode oscillator level can be manually adjusted. No polarization information is available with this receiver and polarization will not be addressed in this article.

We observed the Sun and Moon on 20 January, 2015, which is a New Moon day. The elevations of the Sun and Moon at the local Noon were about 34.2 and 37.0 degrees, respectively. Their total angular separation was about nine degrees. The central regions of the Sun and New Moon were observed quasi-simultaneously using raster scans. For the calibration observation, we scanned the Sun and Moon along the declination direction, with each scan having a length of 7200´´ and 120 second duration. Hence, the scan speed was 60´´ s$^{-1}$. The detected signal was sampled using a 10-Hz analog-to-digital converter (6´´ per sample).

The mapping observation of the Sun was taken just after the calibration observations. We scanned the Sun along the right ascension direction. Each map contained 32 scans separated by 7´´, and each scan had a length of 3600 and 60-second duration (6´´ per sample). This resulted in Nyquist sampling of the mapped region in both the right ascension and declination directions. For calibration purposes, we also measured an OFF point that was more than 3600´´ from the disk center (in order to exclude solar contamination) before beginning the raster scans. Observation of the entire region required a period of 40 minutes. Due to solar rotation, a point initially at solar disk center moves 6.2´´ westward over the term of the observation, which is less than half the beam size of the 45-m telescope at 115 GHz. Hence, the effect of solar rotation can be neglected in this study.

**2.2 Detuning Method**
At millimeter wavelengths, the Sun has brightness temperatures of up to 10,000 K. Such high brightness temperatures exceed the linear-response range of receivers designed for high-sensitivity observations, such as the receiver we used. Therefore, the solar emission must be attenuated in order to prevent saturation of the receiver system.

The Nobeyama 45-m telescope has been used previously to observe the Sun. However, those prior studies did not attenuate the solar emission before it was fed to the receiver systems (Kosugi *et al*., 1986; Irimajiri *et al*., 1995) and, therefore, the linearity of the measurements was not guaranteed. In a recent study, a radio attenuation material (a so-called "solar filter") was used to prevent the receiver system from becoming saturated (Iwai and Shimojo, 2015). Using the solar filter, however, the ratio between the ambient temperature level and the sky level was too small to calibrate the system using the beam-chopper method (Penzias and Burrus, 1973). Instead, the observed signal was normalized using the quiet region and the relative brightness temperatures of the chromospheric structures derived in previous studies.

In this study, we employed the "detuning method" (Yagoubov, 2013) to induce different bias voltages onto the SIS mixer. Figure 1 shows a schematic image of the relationship between the bias voltage and SIS current (SIS DC I-V curves) with two different local (LO) signals and their SIS output levels. SIS mixers are usually tuned through the application of the bias voltage at which the maximum output power is obtained (labeled "Normal" in Figure 1). Further, SIS devices typically have a second peak at a bias voltage value $nhf/e$ lower than that at which the maximum output power is obtained, where $h$ is the Planck constant, $f$ is the LO frequency, $e$ is the elementary charge, and $n$ is the number of series junctions in the device. This second peak has a lower conversion gain and a wider dynamic range but less sensitivity than that of the maximum peak. However, the degradation of the sensitivity is not critical for solar observations because the solar emission is so strong. This tuning condition is referred to as "mixer detuning 1" (MD1) in this study. Another peak exists at the bias voltage $nhf/e$ higher than the maximum output power point, which we call "mixer detuning 2" (MD2).

In this study, we tested a detuning method that induces a lower-level LO signal on the SIS mixer (indicated by the dotted lines in Figure 1). The dynamic range of a SIS mixer can be further improved when we induce a weaker LO signal, even though the sensitivity of the SIS mixer is also degraded. We refer to this tuning condition as mixer detuning 0 (MD0). We can also tune the SIS mixer using both a lower LO level and lower bias voltage (MD01), or a lower LO level and higher bias voltage (MD02).

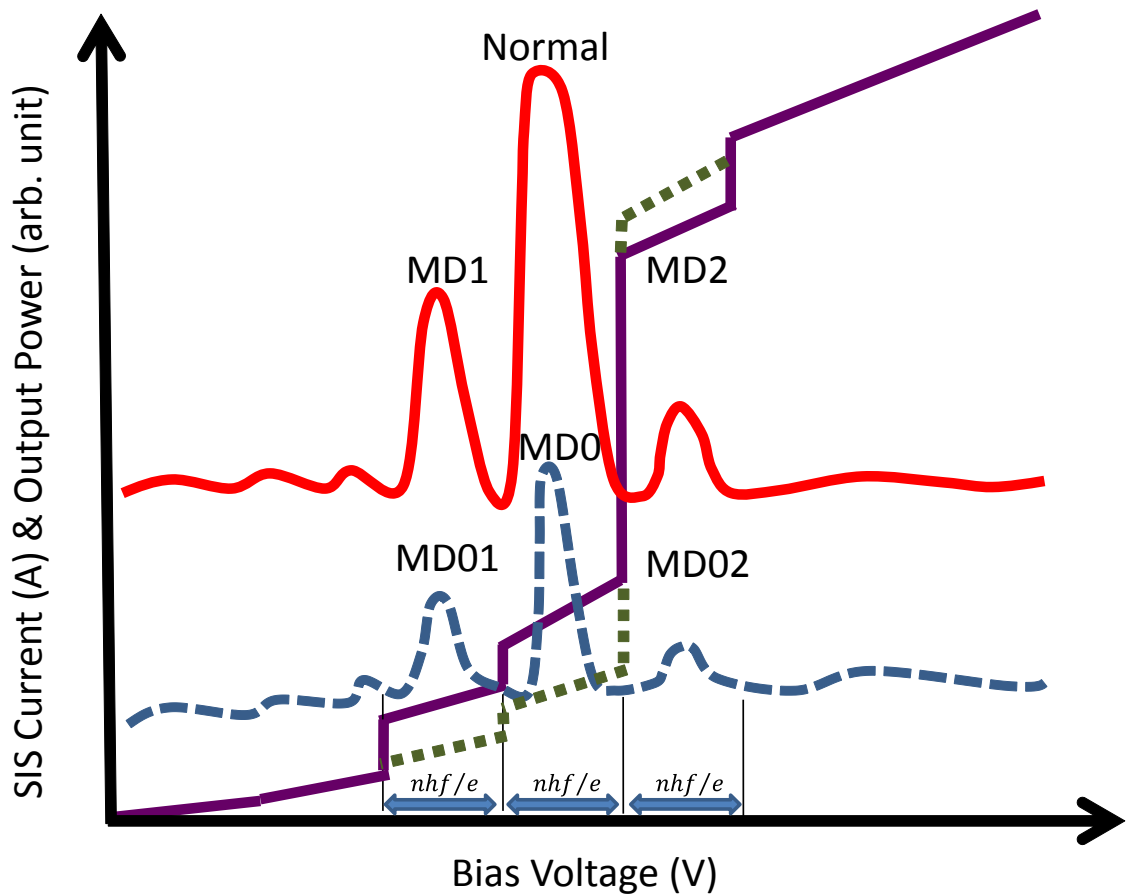

**Figure 1** Schematic image of SIS DC I-V curves with LO signals (purple-solid line: normal LO condition, green-dotted line: lower LO condition), and output diagrams of an SIS mixer (red-solid curve line: normal LO condition, blue-dashed curve line: lower LO condition).

We examined the linearity of the receiver system using the Sun itself. The brightness temperature of the Sun should vary sharply at the solar limb (Ewell *et al*., 1993); however, the broad sidelobes of the antenna response usually produce a gradual change in the observed solar edge (this is explained in more detail below). This effect enables us to recognize that the scan profile of the Sun varies gradually from the sky level to the disk center level (approximately 7000 K). We observed the Sun using different tuning conditions almost simultaneously. Figure 2 shows the scatter plots obtained for two different mixer modes (combinations of Normal, MD01, or MD02 modes), where MD01 and MD02 are tuned at the same local level. The red and green-dashed lines show the results obtained using a signal derived inside the disk region and outside the limb, respectively. If neither of the two tuning modes is saturated, a linear scatter plot should be obtained. However, the scatter plots obtained for the combinations of the Normal,

MD01, and MD02 modes exhibit non-linearity at the solar disk level. This is caused by saturation of the SIS device as a result of the high input level. The calibration load level is typically lower than the solar disk level and should be included in the linearity range of the SIS receiver. Hence, the Normal mode underestimates the solar disk level.

We assumed that the SIS device outputs a lower signal power under "deeper detuned" conditions. For example, if we input the same signal, the output level of the MD02 mode should be lower than that of the MD01 mode. Hence, if two output levels derived under different tuning conditions exhibit linearity, this indicates that neither of these conditions is saturated. In Figure 2, linearity between the MD01 and MD02 modes can be seen within the margin of error. Therefore, this result indicates that neither the MD01 nor the MD02 mode is saturated. We primarily used the MD0 mode, which was confirmed to be non-saturated by the MD01 and MD02 modes in our observation.

The system noise temperature $[T_{\text{sys}}]$ is given by the following equation,

$$T_{\text{sys}} = T_{\text{amb}} \frac{P_{\text{sky}}}{P_{\text{amb}} - P_{\text{sky}}} \quad (1)$$

where $T_{amb}$ is an ambient temperature, and $P_{sky}$ and $P_{amb}$ are output powers of the detector for the blank sky and the beam chopper (assumed to be at the ambient temperature), respectively. In Figure 2, $T_{\text{sys}}$ was 750, 1500, and 2250 for normal, MD01, and MD02, respectively. This result clearly shows that the deeper detuned mode has a larger $T_{\text{sys}}$.

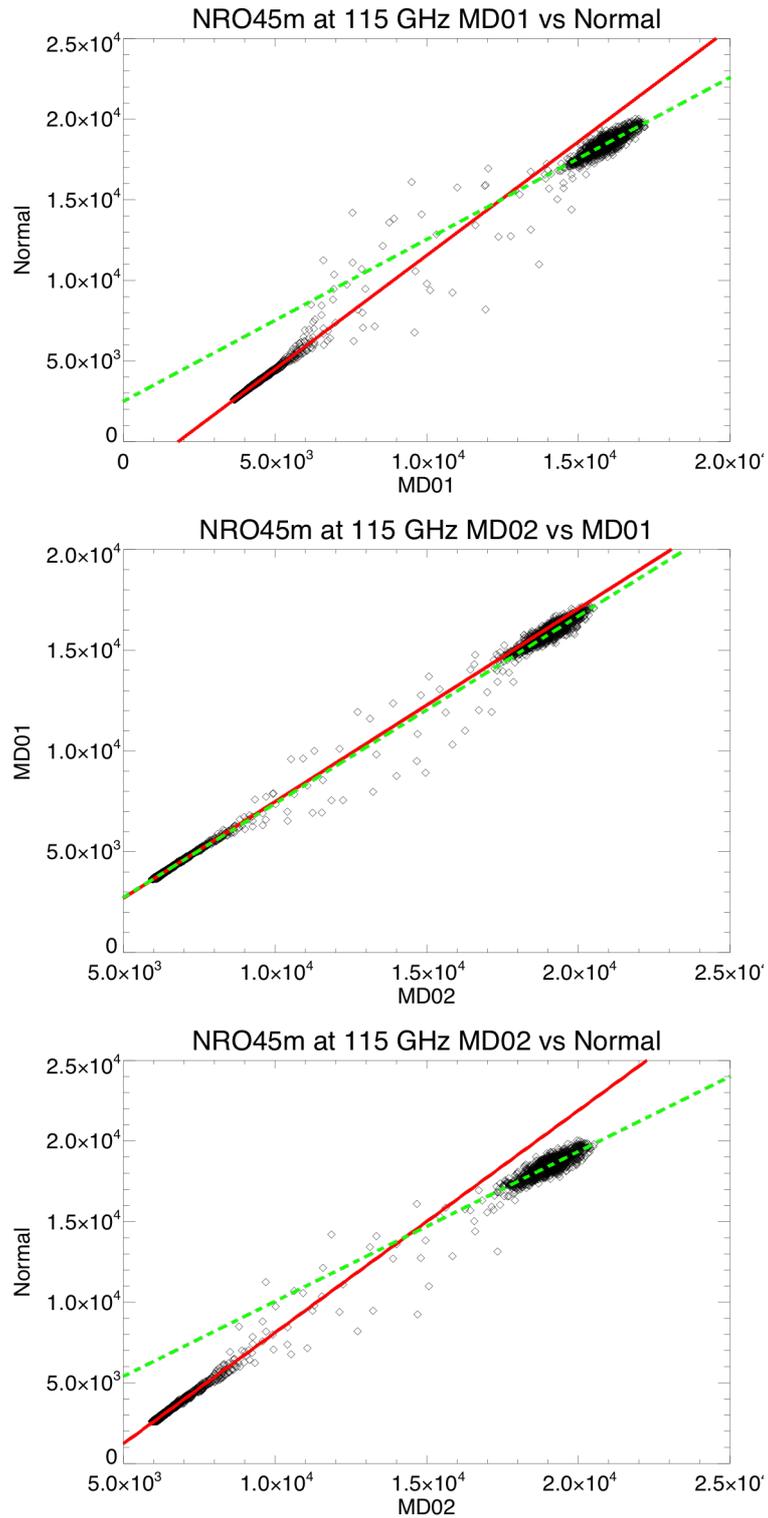

**Figure 2** Scatter plots of solar output levels for two different mixer modes: (top) Normal and MD01; (middle) MD01 and MD02; and (bottom) MD02 and Normal. The red and green-dashed lines show the fitting results obtained using a signal derived inside the disk region and outside the limb, respectively.

**2.3 Calibration Using New Moon**

The Sun is a broad radio source with high brightness, and the majority of the sidelobes of typical millimeter telescopes fall within the solar disk when they point at disk center. Hence, the main-beam efficiency, which excludes the contributions of the sidelobes, should not be used when deriving the real solar brightness temperature from the antenna temperature. The real brightness temperature of the target source $[T_B]$ can be derived as

$$T_B = \frac{T_A^*}{\eta_c \eta_{fss}}, \qquad (2)$$

where $T_A^*$ is the antenna temperature of the target source, $\eta_c$ is the source-coupling efficiency, and $\eta_{fss}$ is the forward spillover and scattering efficiency (Kutner and Ulich, 1981). Note that $\eta_c$ is usually determined by the coupling between the telescope beam and the target source. For the solar and lunar observations, we can assume that all sidelobes are included in the disk when the telescope points near the disk center. In addition, brightness temperature models of the Moon $[T_m]$ have been developed in several studies (Linsky, 1966; 1973a). Therefore, we can approximate the $\eta_c \eta_{fss}$ of a telescope based on the antenna temperature of the Moon $[T_{AM}^*]$, which we define as lunar filled beam efficiency $[\eta_{moon}]$.

$$T_m \approx \frac{T_{AM}^*}{\eta_{moon}} \qquad (3)$$

where $\eta_{moon} \approx \eta_c \eta_{fss}$, which can be measured by lunar observations. The Moon has almost the same visible size as the Sun. Hence, the derived $\eta_{moon}$ can be used to scale the antenna temperature of the Sun to the brightness temperature (Linsky, 1973b; Bastian *et al.*, 1996). The commonly used beam efficiency "main beam efficiency $[\eta_{mb}]$" is given by

$$\eta_{mb} = \eta_c(main\ beam)\eta_{fss} \qquad (4)$$

where $\eta_c(main\ beam)$ is a term that describes the coupling to a source that just fills the main beam, and it can be approximated to be unity at the solar and lunar disk center (Mangum, 1993). However, main beam efficiencies are usually derived from radio sources which are comparable to or smaller than the main beam size of the telescope, which means $\eta_c(main\ beam) \neq 1$. Therefore, they should not be used to calibrate the sources that are much larger than the main beam size, such as the Sun and Moon.

In this study, we derived $T_A^*$ using a single-beam chopper. It is possible to define

$$T_A^* = \frac{T_A \exp(\tau \sec Z)}{\eta}, \qquad (5)$$

where $T_A$ is the observed antenna temperature, $\eta$ is the feed efficiency of the telescope, which includes the rear spillover and scattering efficiency and ohmic loss effect, $\tau$ is the optical depth at the zenith, and $Z$ is the zenith angle. The output powers of the detector for the blank sky $[P_{\text{sky}}]$, the target source $[P_{\text{source}}]$, and the beam chopper having ambient temperature $[P_{\text{amb}}]$ are defined as

$$P_{\text{sky}} = Gk[T_{\text{RX}} + (1-\eta)T_{\text{amb}} + \eta\{1-\exp(-\tau \sec Z)\}T_{\text{atm}}],$$
$$P_{\text{source}} = Gk[T_A + T_{\text{RX}} + (1-\eta)T_{\text{amb}} + \eta\{1-\exp(-\tau \sec Z)\}T_{\text{atm}}], \quad (6)$$
$$P_{\text{amb}} = Gk[T_{RX} + T_{amb}],$$

respectively, where $G$ is the gain of the receiver system, k is the Boltzmann constant, $T_{\text{RX}}$ is the system noise temperature, $T_{\text{amb}}$ is the ambient temperature, and $T_{\text{atm}}$ is the atmospheric temperature. In the above equation, the rearward spillover beams are assumed to be terminated to the ground, which has the same temperature as the ambient atmosphere. The cosmic background radiation is ignored in the above equations. Here, we define the approximate antenna temperature $[T_A']$ as

$$T_A' = T_{\text{amb}} \frac{P_{\text{source}} - P_{\text{sky}}}{P_{\text{amb}} - P_{\text{sky}}} = \frac{T_{\text{amb}} T_A}{\eta[T_{\text{amb}} - \{1-\exp(-\tau \sec Z)\}T_{\text{atm}}]}. \quad (7)$$

Unfortunately, we cannot derive $\tau$ and $T_{\text{atm}}$ because the current Nobeyama 45-m telescope system does not have a water-vapor radiometer, which would enable us to estimate sky conditions. However, if we assume that $T_{\text{atm}} = T_{\text{amb}}$, $T_A'$ becomes $T_A^*$ and we can derive $T_A^*$ from the three observational outputs and $T_{\text{amb}}$. This assumption is discussed further in the latter part of this section.

Figure 3 shows an example of the scan profiles of the Sun and the New Moon. The top panel shows the antenna temperature of the New Moon scan. We derive $T_{\text{AM}}^*$ from the above equations. The $\eta_{\text{moon}}$ of the telescope is defined as

$$\eta_{\text{moon}} = \frac{T_{AM}^*}{T_m}, \quad (8)$$

where $T_m$ is the modeled brightness temperature of the Moon. In this study, we used the lunar brightness temperature model proposed by Linsky (1973a). We measured the $\eta_{\text{moon}}$ of the 45-m telescope at 115 GHz using the New Moon, obtaining a result of approximately 0.74.

The antenna temperature of the Sun $[T_{\text{AS}}^*]$ is also defined by Equation (7). The middle panel of Figure 3 shows the antenna temperature of the solar scan. The brightness temperature of the Sun $[T_{\text{BS}}]$ can be expressed as

$$T_{\mathrm{BS}} = \frac{T_{\mathrm{AS}}^*}{\eta_{\mathrm{moon}}}. \qquad (9)$$

The bottom panel of Figure 3 shows the absolute brightness temperature of the solar scan, which is calibrated using $\eta_{\mathrm{moon}}$. Note that both the Sun and Moon should have a sharp limb. However, the observed scan profiles exhibit a gradual change because the broad sidelobes of the telescope response affect the scan profile. This effect is discussed in Section 2.4

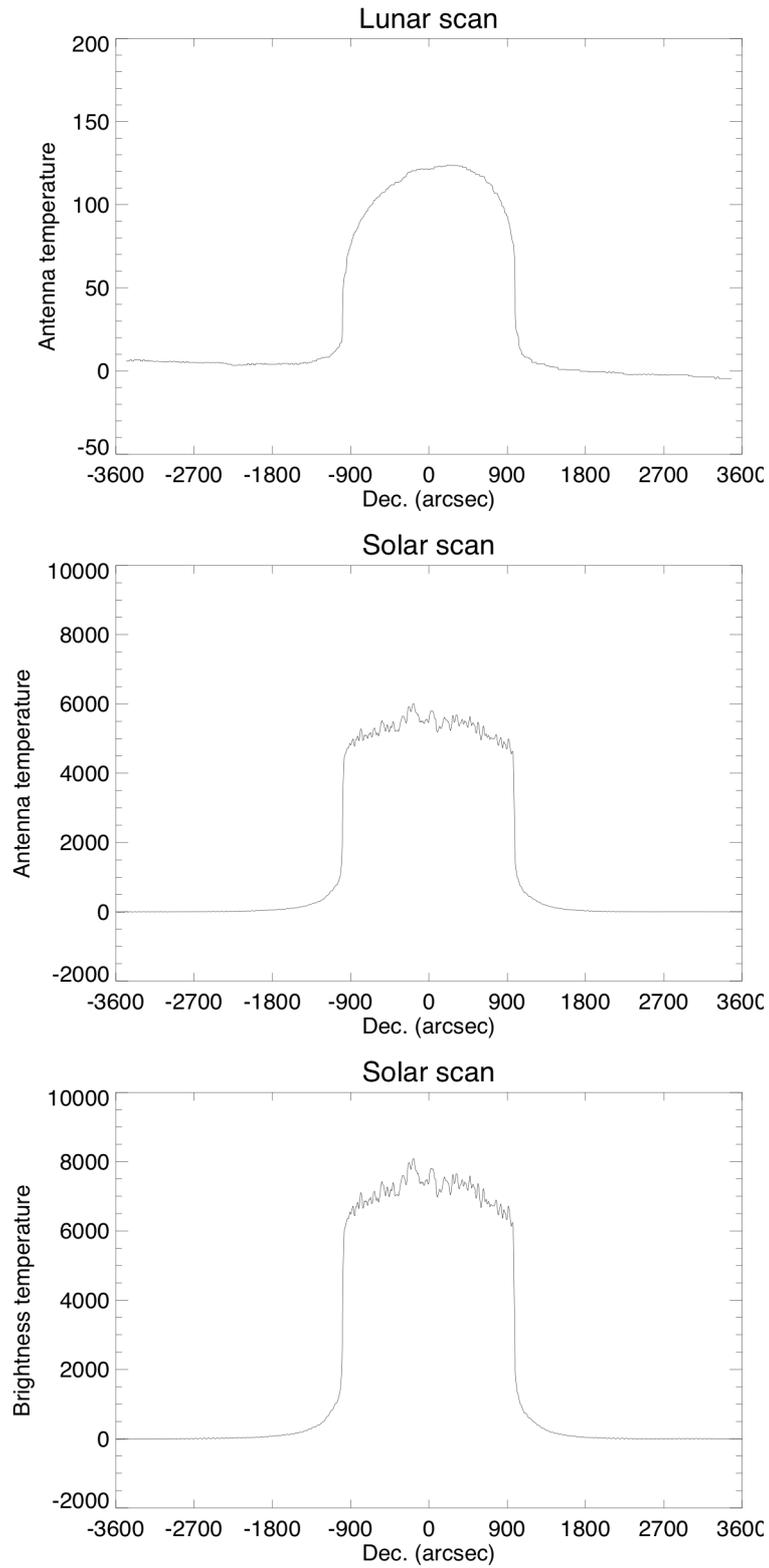

**Figure 3** Scan profiles of antenna temperatures of (top) the Moon, and (middle) the Sun. (bottom) Scan profile of the Sun, calibrated to brightness temperature.

As regards single-load calibration, we typically require an assumption that the ambient temperature is equal to the sky temperature ($T_{atm} = T_{amb}$ in Equation (7)). This assumption is usually not true. Hence, the approximated antenna temperatures contain some error ($T_A^{'} \neq T_A^*$ in Equation (7)). The difference between the ambient temperature and atmospheric temperature is probably about 20 K for typical sky conditions at the observatory. For example, if we assume $T_{amb}$, $T_{atm}$, and $\tau \sec Z$ to be 273, 253, and 1, respectively, the difference between $T_{AM}^{'}$ and $T_{AM}^*$ should then be approximately 10 %, which can be recognized as the typical error of $T_{AM}^*$ [$\sigma_{T_{AM}^*}$]. The error in the lunar brightness temperature [$\sigma_{T_m}$] has been estimated to be approximately 4 % (Linsky, 1973a). The error included in $\eta_{moon}$ [$\sigma_{\eta_{moon}}$] can be estimated as follows,

$$\sigma_{\eta_{moon}}^2 = \left(\frac{\partial \eta_{moon}}{\partial T_{AM}^*}\right)^2 \sigma_{T_{AM}^*}^2 + \left(\frac{\partial \eta_{moon}}{\partial T_m}\right)^2 \sigma_{T_m}^2 \quad (10)$$

The estimated $\sigma_{\eta_{moon}}$ is 0.08 (about 11% of the derived $\eta_{moon}$). Thus we find $\eta_{moon}$ at 115 GHz is 0.74 ± 0.08.

**2.4 Beam Pattern Estimation using Solar Limb**

Larger millimeter and sub-millimeter telescope reflectors are usually composed of many panels, whose edges typically create weak but broad sidelobes. The beam patterns and antenna efficiencies of these radio telescopes are usually measured using planets and the Moon (Mangum, 1993). In fact, the Moon has been used to estimate the sidelobes of many telescopes, such as in the case of the Insitut de Radioastronomie Millimétrique (IRAM) by Greve *et al*., (1998), Atacama Sub-millimeter Telescope Experiment (ASTE) by Sugimoto *et al*., (2004), and Atacama Large Millimeter/Sub-millimeter Array (ALMA) by Sugimoto *et al*., (2009). The Sun is a broad radio source with high brightness, and even weak sidelobes can affect the observational results if not accounted for. Therefore, the Sun itself is sometimes used to estimate broad sidelobes, as in the case of the Caltech Sub-millimeter Observatory (CSO) by Bastian *et al*. (1993) and the James Clerk Maxwell Telescope (JCMT) by Lindsey and Roellig (1991).

The solar limb is known to appear as a sharp edge at millimeter wavelengths, based on eclipse observations (*e.g.*, Lindsey *et al*., 1992; Ewell *et al*., 1993). However, scan profiles from single-dish observations typically exhibit a gradual change, because millimetric telescopes have broad sidelobes. The main reflector of the Nobeyama 45-m telescope consists of approximately 600 carbon-fiber reinforced plastic (CFPR) panels. The typical panel size is approximately 2 m. Hence, the 45-m telescope has sidelobes that roughly correspond to this panel size, which is more than 20 times broader than the main beam.

These sidelobes usually have an insignificant effect on compact non-solar sources, because they are significantly weaker than the main beam. However, the Sun is a bright radio source that is broader than the majority of the sidelobes. Hence, the effects of the sidelobes are proportional to their area. For example, a sidelobe that is ten-times broader than the main beam has a 100-fold greater effect than the main beam. Therefore, even if that sidelobe is 20-dB weaker than the main beam, it can have the same effect as the main beam.

In this study, a model of the telescope sidelobes was created using the Sun itself. Figure 4 shows the modeling sequences employed. We regarded the beam pattern of the 45-m telescope as being a combination of three Gaussian functions (the main beam, a relatively narrow sidelobe, and a broad sidelobe; Figure 4a), and the beam model equation (the point-spread function (PSF) for the deconvolution) was expressed as

$$f(\nu) = \exp\left(-r^2 \left(\frac{2\sqrt{\ln 2}}{15"}\right)^2\right) + a \exp\left(-r^2 \left(\frac{2\sqrt{\ln 2}}{b"}\right)^2\right) + c \exp\left(-r^2 \left(\frac{2\sqrt{\ln 2}}{d"}\right)^2\right), \quad (11)$$

where $\nu$ is the frequency and $r$ is the radial distance from the beam center ($r^2 = x^2 + y^2$). This PSF contains four free parameters ($a - d$). The Sun was regarded as a disk with a sharp edge (Figure 4b). We assumed that the disk-center region was not affected by the sidelobes. Therefore, the brightness temperature of the quiet region around the disk center, which was derived in Section 2.2, was used as the brightness temperature of the modeled Sun. We performed a convolution of the 45-m telescope beam with the modeled Sun (Figure 4c), which was expected to conform to the observed scan profile. Further, iteration of the four free parameters of the two sidelobes to minimize the difference between the observed and convolved limb profiles was expected to provide the optimum sidelobe value. Figure 4d shows center slices of the modeled and observational solar disks.

The sidelobe parameters derived in this study are summarized in Table 1. It should be noted that these sidelobes were derived under solar illumination of the dish surface, and Ukita (1998) has reported that thermal deformations of CFPR panels occur as a result of exposure to sunshine. Hence, the derived sidelobe model should not be used for non-solar sources. We should also mention that the sunshine and focusing conditions are not always identical. Hence, it is not surprising that the derived sidelobe parameters differ depending on the observations.

**Table 1** Sidelobe parameters of 45-m telescope at 115 GHz.

| Observing day | $a$ | $b$ [arcsec] | $c$ | $d$ [arcsec] |
|---|---|---|---|---|
| January 20, 2015 | 1.25 | 55 | 0.0040 | 710 |

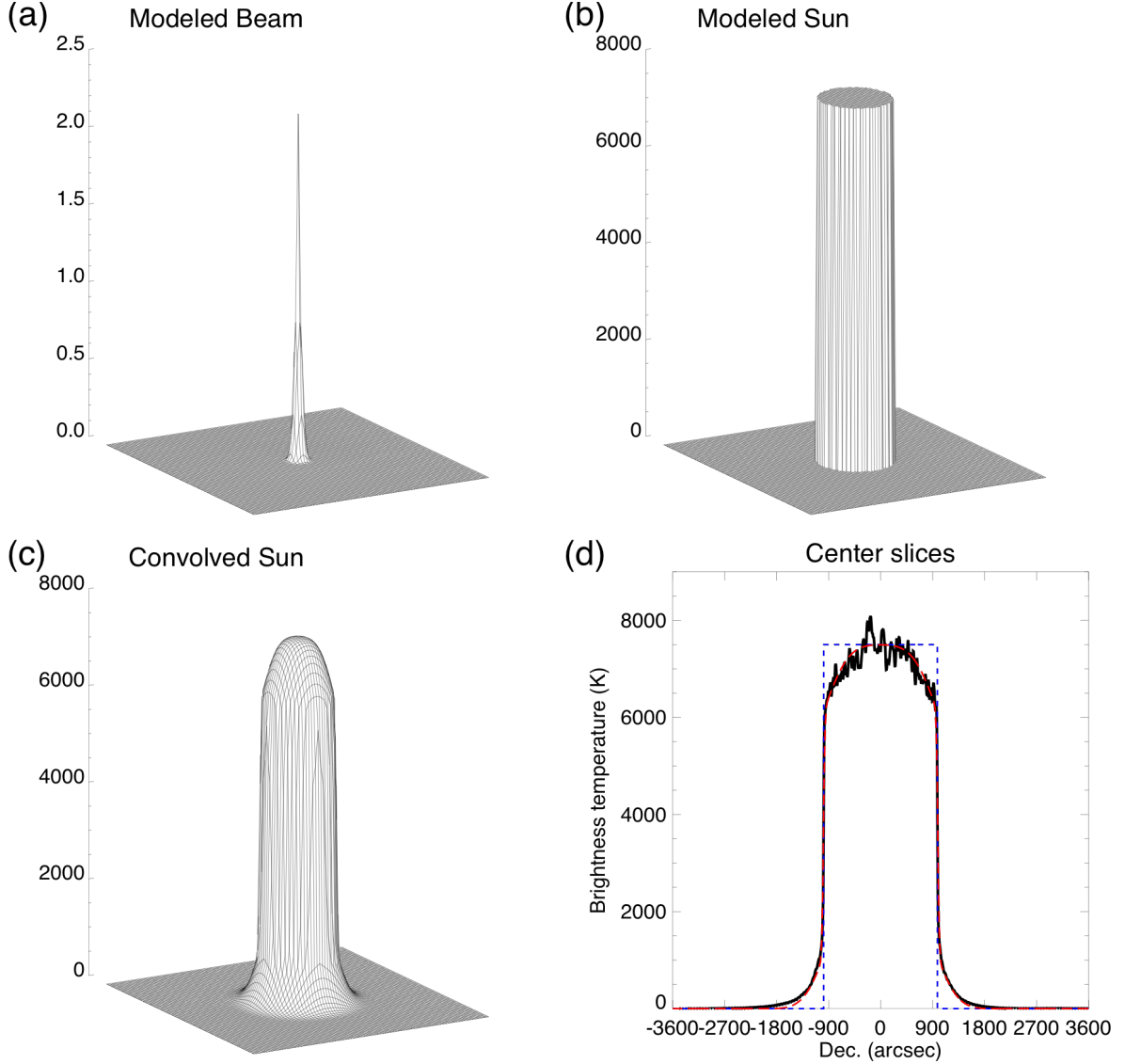

**Figure 4** (a) Modeled beam pattern of 45-m telescope at 115 GHz. (b) Solar-disk model with sharp edge at limb. (c) Solar-disk model convolved with 45-m telescope beam pattern. (d) Scan profiles of solar-disk center. Black, blue-dashed, and red-long-dashed: Observational result, solar-disk model, and solar-disk model convolved with 45-m telescope beam pattern, respectively.

## 2.5 Side-lobe Deconvolution

In this study, we derived the actual solar disk map using the 2D deconvolution of the telescope beam pattern from the observed map using the maximum likelihood method (`max_likelihood.pro` in SSWIDL). The 45-m telescope can scan only a limited region of the solar disk within the time frame that allows us to ignore the solar rotation and atmospheric variations. Hence, we used the solar-disk model convolved with the 45-m telescope beam pattern (shown in Figure 4c) to complement the remainder of the scanned disk region. Figure 5 shows the deconvolved solar disks. Each solar disk contains a scan region near the center of the disk that demonstrates the variation in $T_{BS}$ due to the active regions and plages. The remainder of the disk region, which exhibits a smooth shape, is the complement model.

The appropriate number of iterations prior to termination of the deconvolution processes is determined by the limb shape. Figures 5a, 5b, and 5c show the solar disks obtained after 3, 10, and 30 iterations, respectively. The center slices of each scan are given in Figure 5d. The original scan profile (black) has a broad wing outside the limb regions. The wing of the limb becomes smaller and the limb shape becomes a sharp edge as the number of iterations increases up to ten (green). However, after the iterations have been performed more than ten times, no further significant changes in the limb shape occur (see the red profile in Figure 5d) which is derived after 30 iterations). On the other hand, the disk region immediately inside the limb achieves an anomalously high brightness temperature as a result of numerical divergence. Hence, the iteration number was set to ten for the remainder of this study.

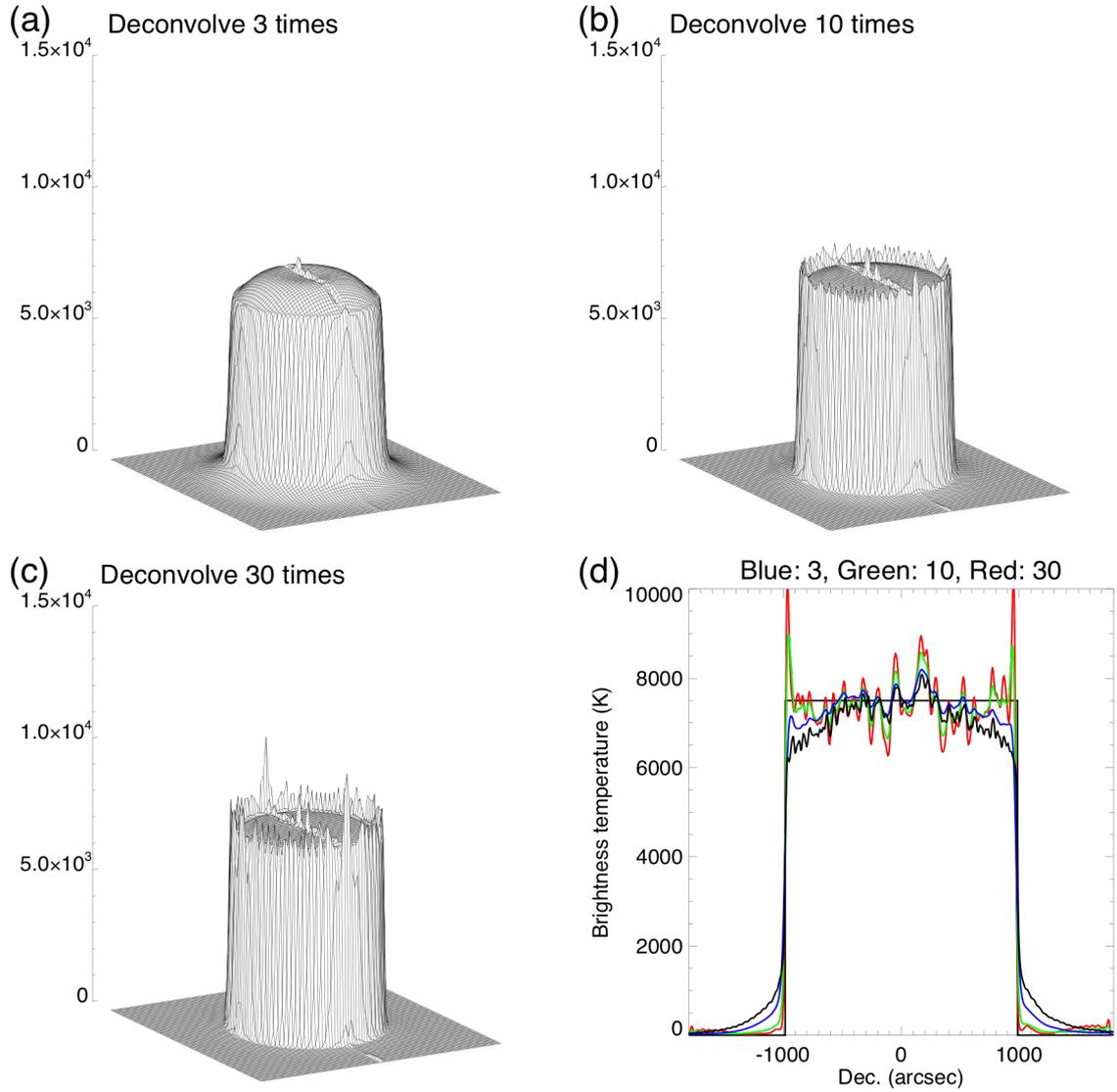

**Figure 5** Solar disks following (a) 3, (b) 10, and (c) 30 deconvolutions. (d) Scan profiles of the centers of the obtained solar disks. Black, blue, green, and red: Original observational result and results following 3, 10, and 30 deconvolutions, respectively.

In this study, we used the solar-disk model convolved with the 45-m telescope beam pattern to complement the remainder of the scanned disk region. However, the actual solar disk contains various chromospheric structures. Hence, it should be mentioned that the beam spills over onto the modeled disk region and this may result in errors in the deconvolved map. The modeled sidelobes in this study have two sizes: the narrower sideobe is about 55˝ and the other is about 700˝. The region within 55˝ from the boundary between the model and observed regions can be affected by the error from the spillover of the narrow sidelobes. The large sidelobe also can affect the result. However, most chromospheric structures are smaller than the spatial scale of the large sidelobe. Therefore, the error from the coupling between the large sidelobe and the brightness temperature perturbations from the chromospheric structures is small. The error from the coupling between large sidelobe and disk scale structures should also be addressed. However, the large-scale disk shape should have axial symmetry, and the modeled region should be similar to the actual structures. Therefore, the limb structures in the deconvolved map do not have significant error from the spillover sidelobes.

## 3. Results and Discussion
### 3. 1 Brightness Temperature of the Sun

We assume that the observed radio emission is pure thermal free–free emission which, at 100 GHz, is dominated by opacity from free electrons colliding with ions (*e.g.* Wedemeyer *et al*., 2016). Gyro-resonance emission, in which opacity is provided by electrons spiraling in magnetic fields under the Lorentz force, also contributes at centimeter wavelengths, but it would require field strengths in the atmosphere of order 10,000 G to contribute at 100 GHz (*e.g.*, White, 2004) and so can be ignored here.

The solar disk contains various structures such as active regions, plages, coronal holes, prominences, and network structures. Hence, definitions of the "quiet region" can vary. In this study, we defined the quiet region as the region that contains neither sunspots nor plages, using a UV image at 1700 Å obtained by the *Atmospheric Imaging Assembly* (AIA: Lemen *et al*. 2012) onboard the *Solar Dynamics Observatory* (SDO). We also considered the center region of the solar disk only, so as to avoid the sidelobe effect described in Section 2.4. Therefore, we employed the observational results before the sidelobe deconvolution. Figure 6 shows the region observed on 20 January, 2015 by the raster scans conducted in this study. The red rectangle in Figure 6a indicates the quiet region considered here. We then produced a histogram of the brightness temperature of the quiet region, which was fit in accordance with a Gaussian function. The center of the Gaussian function was defined as the quiet-Sun level. The brightness temperature of the quiet Sun obtained in this study at 115 GHz is 7700 K.

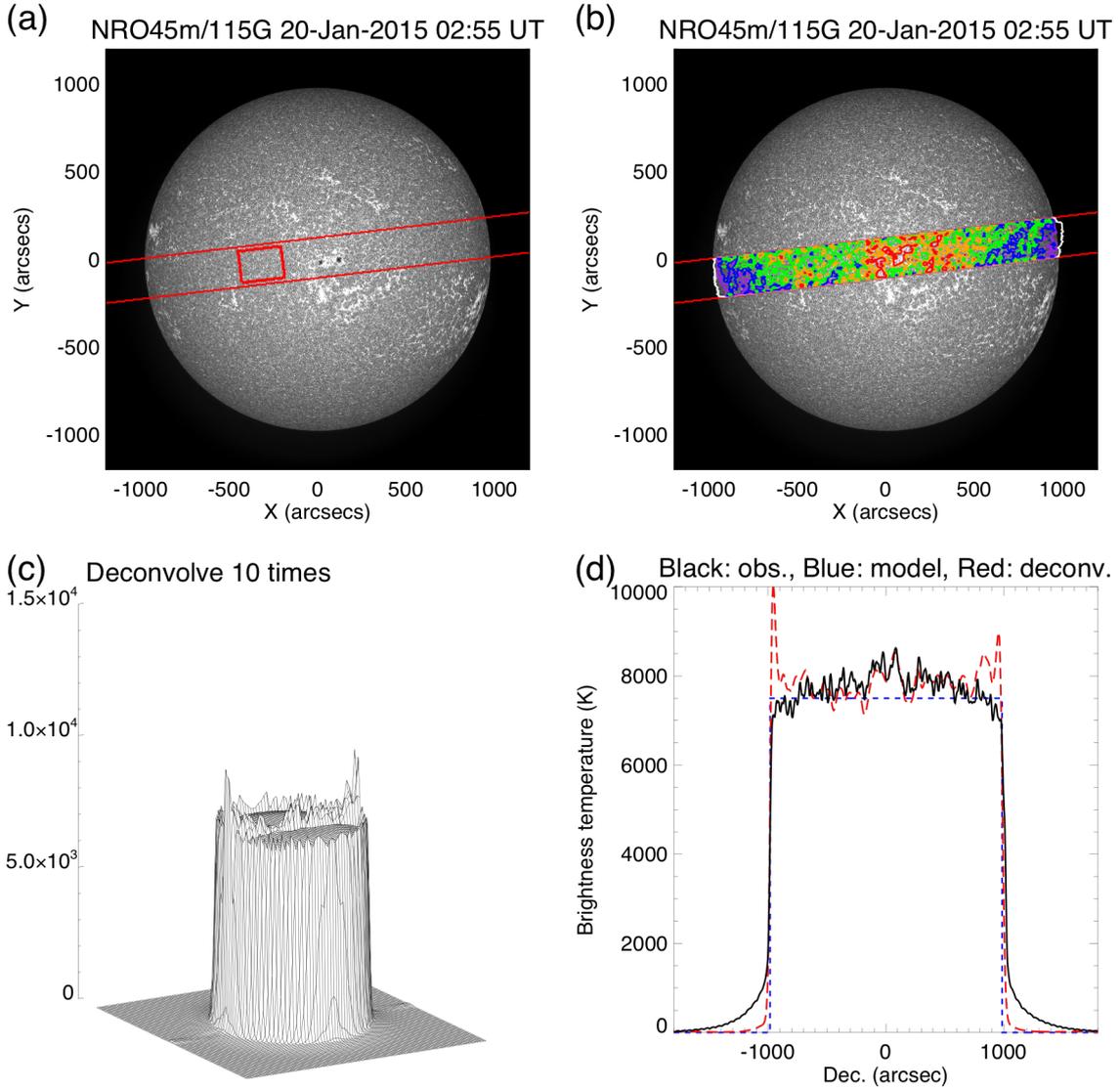

**Figure 6** (a) Full solar-disk image at 1700 Å observed on 20 January, 2015. (b) Radio contour map observed at 115 GHz without sidelobe deconvolution overlaid on 1700-Å image. Red, orange, green, blue, and purple: 8300, 8000, 7700 (quiet region level), 7400, and 7100 K, respectively. (c) Solar disk after ten iterations of the sidelobe deconvolution. (d) Scan profiles of the center of the solar disk. Black, blue-dashed, and red-long-dashed: Observational result, solar disk model, and observational result deconvolved with the 45-m telescope beam pattern, respectively.

### 3.2 Error Estimation

Various sources contribute to the uncertainty in the $T_{BS}$ results. The Sun, Moon, and ambient load are all strong radio sources, and Figure 3 indicates that sufficient signal-to-noise ratios for analysis are obtained under the imposed detuning conditions. The linearity of the instrument under solar observations was confirmed using the data given in Figure 2. If part of a sidelobe extends outside the lunar disk, $\eta_{\text{fss}}$ may be underestimated. However, the derived sidelobe model shown in Figure 4 suggests that the broad sidelobe of the Nobeyama 45-m telescope is smaller than the solar and lunar disks. Therefore, the above factors do not constitute a significant source of error in our observation. Further, it is possible that the beam shape and efficiency may differ between the solar and lunar observations, but it is difficult to obtain a quantitative estimation of this difference. Such differences are likely minimized in this study through use of the observation of the New Moon quasi-simultaneously (all observations were conducted within a period of 1 hour) with the observation of the nearby Sun. Hence, this effect should also be a minor source of error, because antenna deformation typically depends on the elevation angle and sunlight conditions.

However, if we assume that all atmospheric conditions ($T_{\text{atm}}$, $T_{\text{amb}}$, and $\tau \sec Z$) are identical for the solar and lunar observations, the difference between $T_A^{'}$ and $T_A^{*}$ should also be identical for the Sun and Moon, such that

$$T_{\text{AS}}^{'} = CT_{\text{AS}}^{*},$$
$$T_{\text{AM}}^{'} = CT_{\text{AM}}^{*}, \quad (12)$$

where $T_{\text{AS}}^{'}$ and $T_{\text{AM}}^{'}$ are the approximate antenna temperatures of the Sun and Moon, respectively, and $C$ is a constant. This assumption and equations (8) and (9) indicate that

$$T_{\text{BS}} = \frac{T_{\text{AS}}^{*}}{\eta_{\text{moon}}} = \frac{T_{\text{AS}}^{*}}{T_{\text{AM}}^{*}} T_m = \frac{T_{\text{AS}}^{'}}{T_{\text{AM}}^{'}} T_m. \quad (13)$$

Hence, we can derive $T_{\text{BS}}$ regardless of the atmospheric conditions and $\eta_{\text{moon}}$. This is the greatest advantage to using the New Moon as a solar calibrator. Therefore, the primary source of error in Equation (11) is the error in the lunar brightness temperature, which has been estimated to be approximately 4 % (Linsky, 1973a). Hence, we assume that the minimum limit of the error in the solar brightness temperature in this study is $\pm 4$ %. Hence, the $T_{\text{BS}}$ at 115 GHz derived in this study is 7700 $\pm$ 310 K.

The above assumption is correct only for a restricted scenario in which the Sun and Moon are in the same location and are observed simultaneously. These conditions are met only in the case of a solar eclipse. On the other hand, our observations were performed under New Moon conditions. Although the Sun and Moon are positioned more closely together on a New Moon day, the atmospheric conditions from the telescope to the Sun and Moon should not be assumed to be equal. Therefore, our results contain a greater degree of error than indicated above, which is difficult to estimate. This is a limitation of the one-temperature calibration approach. The worst case can be estimated to assume that the entire atmospheric conditions toward the Sun and Moon are not canceled in the calculation of $T_{BS}$. The error $\sigma_{T_{BS}}$ can be estimated

$$\sigma_{T_{BS}}^2 = \left(\frac{\partial T_{BS}}{\partial T_{AM}^*}\right)^2 \sigma_{T_{AM}^*}^2 + \left(\frac{\partial T_{BS}}{\partial T_{AS}^*}\right)^2 \sigma_{T_{AS}^*}^2 + \left(\frac{\partial T_{BS}}{\partial T_M}\right)^2 \sigma_{T_M}^2 \quad (14)$$

where $\sigma_{T_{AM}^*}$ $\sigma_{T_{AS}^*}$, and $\sigma_{T_M}$ are the uncertainties in $T_{AM}^*$, $T_{AS}^*$, and $T_M$, respectively. We assume that $\sigma_{T_{AM}^*}$, $\sigma_{T_{AS}^*}$, and $\sigma_{T_M}$ are 10 %, 10 %, and 4 %, respectively, as estimated in Section 2.3. We derived $T_{AM}^*$ and $T_{AS}^*$ from the top and middle panels of Figure 3, respectively. The worst case of $\sigma_{T_{BS}}$ is derived to be 15 %. In that case, $T_{BS}$ at 115 GHz should be 7700 $\pm$ 1130 K.

Our calibration method relies on the brightness-temperature model of the Moon. If we can simulate all of the beam efficiencies included in the optics, we might be able to derive the absolute brightness temperature without the lunar-brightness-temperature model. It should be mentioned that the further estimations will also require the measurement of the atmospheric opacity. A two-load calibration scheme (*i.e.*, two different calibration temperatures) would enable us to derive $T_{BS}$ more accurately.

In previous studies, where the brightness temperature of the quiet Sun was also derived and recalibrated using New Moon data, the typical reported value for the quiet Sun at the 3-mm range was 7000 – 8000 K (see the summary by Linsky (1973b)). Thus, our result is consistent with the findings of these previous studies within the given margin of error. It should be mentioned that the definition of the "quiet region" can affect the brightness temperature. Further, unlike the present study, previous experiments did not determine the actual quiet region through consideration of optical or UV images. On the other hand, Iwai and Shimojo (2015) have shown that a close relationship exists between the radio brightness temperature and chromospheric structures, which is supported by the findings of this study (see Figure 6b). Hence, it is possible that the divergence of the brightness temperature values derived in the previous studies is caused, in part, by the chromospheric structures contained within each "quiet region."

### 3.3 Limb Brightening

It is generally expected that the chromospheric emission at the solar limb should be brighter than at disk center due to the emitting region being higher in the solar atmosphere; this is known as limb brightening (*e.g.*, Selhorst *et al.*, 2003 and references therein). On the other hand, this study derived the telescope beam pattern from the observed solar disk by assuming a model that was composed of a flat disk with a sharp edge (see Figure 4b). Then, the real solar disk was derived via deconvolution of the derived beam pattern from the observed solar disk. Hence, the deconvolved disk should be flat, regardless of the occurrence of limb brightening. In fact, the deconvolved polar scan in Figure 5d shows a flat disk with a sharp edge, suggesting that the deconvolution conducted in this study was successful. However, the deconvolved equatorial scans in Figures 6c and 6d exhibit a brighter limb, thus appearing to indicate limb brightening. Iwai and Shimojo (2015) have found that the brightness of the millimetric radio emission corresponds to the chromospheric structure. This good correspondence was observed not only for prominent structures, such as active regions and plages, but also in fine structures, such as the network structures of the magnetic fields. In this article, several network magnetic fields are apparent in the limb regions of Figure 6a. As the radio emission from these structures can cause brighter limbs, this may have led to pseudo limb brightening. Note that the real limb brightening should exist at the millimeter wavelengths. However, it cannot be deconvolved from the beam that derived from the flat disk model.

Some previous studies have suggested limb brightening due to deconvolution of the sidelobes, even though the sidelobes in question were derived using the flat-disk model with sharp limbs (*e.g.*, Bastian *et al*., 1993). It is possible that the extent of the limb brightening was overestimated in those studies. It may be possible to obtain an accurate estimation of limb brightening behavior via sidelobe deconvolution with the sidelobes being derived using a solar disk model incorporating accurate limb brightening. However, this is difficult without accurate limb brightening observations. Therefore, a good radio telescope with a sidelobe level that is significantly smaller than that of the 45-m telescope should be employed. We should also mention that the limb brightening should be estimated at the polar region, because the equatorial region typically contains a greater number of magnetic structures than the polar region. However, the limb brightening of the polar region varies according to the phases of the solar cycle (Gopalswamy *et al*., 2012). Therefore, a full-disk scan would be more effective as regards defining the effect of the chromospheric structures on the limb brightening. It should be mentioned that the Moon can also be used to estimate the beam pattern of the telescope.

## 4. Summary
We observed the Sun using the Nobeyama 45-m radio telescope at 115 GHz. The Sun is an extremely strong radio source at millimeter wavelengths. However, accurate measurement of the brightness temperature of the Sun requires performance evaluation of the telescope under extreme conditions; such investigations are not typically performed for non-solar observations. This study used the Sun itself and the New Moon to evaluate the performance of the 45-m telescope, and it enabled solar observation at the 3-mm range. The results are summarized as follows:

- We detuned the SIS receiver by inducing different bias voltages onto the SIS mixer. Then, we examined the linearity of the receiver system using the Sun by comparing two outputs derived from different tuning conditions. Our results show that the MD01 and MD02 modes used in this study were not saturated during solar observation.
- We measured the lunar filled beam efficiency of the 45-m telescope using the New Moon, and then derived the brightness temperature of the Sun. The filled-beam efficiency of the 45-m telescope at 115 GHz is $0.74 \pm 0.08$ and the derived solar brightness temperature is $7700 \pm 310$ K, which is consistent with previous observations.
- The beam pattern of the 45-m telescope was modeled as a summation of three Gaussian functions (main beam, narrow sidelobe, and broad sidelobe). Then, the best model was derived using the solar limb. The derived beam pattern was used to determine the real shape of the Sun through deconvolution of the beam pattern from an observed map.

In ALMA Observing Cycle 4, which will begin in 2016, ALMA will carry out single-dish solar observations. Although the spatial resolution of the ALMA single-dish will be lower than that of the 45-m telescope at the 3-mm range, the fine structures can be derived via interferometer observation, and single-dish observations can be used to derive the absolute brightness temperatures of the diffuse structures. Such observations will be performed using 12-m antennas, which yield beam patterns that contain significantly smaller sidelobes than those of the 45-m telescope. In addition, the ALMA system facilitates a two-load calibration sequence. Furthermore, ALMA will have the ability to observe the full solar disk within ten min at the 3-mm band. Hence, both equatorial and polar regions will be deconvolved on the same map. This means that the ALMA observations will overcome some of the problems discussed in this study. Solar observations using ALMA will facilitate important progress in not only interferometry, but also single-dish observations of the Sun. This will provide us with the opportunity to begin a new stage of solar radio astronomy research.

**Acknowledgements**

We received considerable advice from many members of the Solar ALMA development team (Principal Investigator: Timothy Bastian), for which we are grateful. We also thank Masahiro Sugimoto for valuable discussions. The 45-m radio telescope is operated by Nobeyama Radio Observatory, a branch of the National Astronomical Observatory of Japan. We wish to thank the 45-m telescope team for their support during our observations. The SDO data are courtesy of the National Aeronautics and Space Administration (NASA)/SDO, as well as the AIA science teams. K. Iwai was supported by the Japan Society for the Promotion of Science (JSPS) Research Fellowships.

**Disclosure of Potential Conflicts of Interest** The authors declare that they have no conflicts of interest.